\documentclass[12pt]{article}
\usepackage{amsmath}
\usepackage{graphicx}
\usepackage{lscape}
\usepackage{psfrag,epsf}
\usepackage{enumerate}
\usepackage{natbib}
\usepackage{amssymb,amsfonts,multirow,multicol,hyperref,bbm,fullpage}
\usepackage{bmpsize}
\usepackage{graphics}
\usepackage{booktabs}
\usepackage{setspace}
\usepackage{theorem}
\usepackage{geometry}
\usepackage{array}
\usepackage{bm}
\usepackage{lscape}
\usepackage{pdflscape}
\usepackage{times}

\usepackage{longtable}
\usepackage{rotating}
\usepackage{hyperref}
\usepackage{xcolor}
\usepackage{ulem}
\usepackage{color}

\usepackage{url} % not crucial - just used below for the URL 

\newcommand{\blind}{0}

\begin{document}

\def\spacingset#1{\renewcommand{\baselinestretch}%
{#1}\small\normalsize} \spacingset{1}

%%%%%%%%%%%%%%%%%%%%%%%%%%%%%%%%%%%%%%%%%%%%%%%%%%%%%%%%%%%%%%%%%%%%%%%%%%%%%%

\if0\blind
{
  \title{\bf The state of play of reproducibility in Statistics: an empirical analysis}
  \author{Xin Xiong\thanks{X. Xiong was supported by a China Council Scholarship}\hspace{.2cm}\\
    Department of Biostatistics, \\Harvard T. H. Chan School of Public Health\\
    and \\
    Ivor Cribben\thanks{
    I. Cribben was supported by the Natural Sciences and Engineering Research Council (Canada) grant RGPIN-2018-06638 and the Xerox Faculty Fellowship, Alberta School of Business.} \\
    Department of Accounting and Business Analytics, Alberta School of Business, \\
    University of Alberta}
  \maketitle
} \fi

\if1\blind
{
  \bigskip
  \bigskip
  \bigskip
  \begin{center}
    {\LARGE\bf Title}
\end{center}
  \medskip
} \fi

\bigskip
\begin{abstract}
Reproducibility, the ability to reproduce the results of published papers or studies using their computer code and data, is a cornerstone of reliable scientific methodology.  Studies where results cannot be reproduced by the scientific community should be treated with caution.  Over the past decade, the importance of reproducible research has been frequently stressed in a wide range of scientific journals such as \textit{Nature} and \textit{Science} and international magazines such as \textit{The Economist}. However, multiple studies have demonstrated that scientific results are often not reproducible across research areas such as psychology and medicine.  Statistics, the science concerned with developing and studying methods for collecting, analyzing, interpreting and presenting empirical data, prides itself on its openness when it comes to sharing both computer code and data.  In this paper, we examine reproducibility in the field of statistics by attempting to reproduce the results in 93 published papers in prominent journals utilizing functional magnetic resonance imaging (fMRI) data during the 2010-2021 period.  Overall, from both the computer code and the data perspective, among all the 93 examined papers, we could only reproduce the results in 14 (15.1\%) papers, that is, the papers provide both executable computer code (or software) with the real fMRI data, and our results matched the results in the paper.  Finally, we conclude with some author-specific and journal-specific recommendations to improve the research reproducibility in statistics.
\end{abstract}

\noindent%
\vfill

\newpage
\spacingset{1.45} % DON'T change the spacing!

\bigskip

%------------------------------------------------------------------------------------------------
\section{Introduction}
Reproducibility and replicability are often cited as the cornerstones of reliable science.  Studies where results cannot be reproduced or replicated by the scientific community should be treated with caution.   The importance and benefits of reproducible and replicable research is well known \citep{donoho,peng2011}.  For authors, the possible benefits include escalating the impact of their research.  This can be achieved because by providing computer code, other researchers can easily use and compare the method developed, which may lead to more citations.  In addition, other possible benefits include elevating work efficiency, improving work habits, communication and teamwork, and minimizing the errors in the research/computer code.  For example, the training of students of all levels becomes dramatically easier, as students can immediately pick up where the previous student left off because the code was written with a reproducible mindset.  Further, the access to the computer code and data enable downstream scientific contributions, such as meta-analyses.  For readers, the benefits include increased trustworthiness, a perceived quality of research, and easy adaption and extension of the computer code and analysis.  For the public, the benefits include reducing or preventing fraud and scandals related to the research, and increasing the public access to the research (public goods).  The quality of the education system can also be improved by encouraging students to interact with the research papers by repeating a part of (or the entirety of) the analysis, rather than being a passive researcher. This practice enriches the experience, creates awareness, and becomes normal practice after they graduate. 

Recently the importance of reproducibility and replicability in research has been frequently stressed in a wide range of scientific journals and magazines, and the terms ``reproducibility crisis'' and ``replication crisis'' have become more evident.  A poll conducted by the journal \textit{Nature} in 2016 reported that more than half (52\%) of scientists surveyed believed science was facing a ``replication crisis'' \citep{baker}.  The crisis involves the absence of replication studies in the published literature across many fields \citep{makel2012} (for example, the Open Science Collaboration found that fewer than half of the studies were successfully replicated in Psychology), the ``file drawer effect'' or the inflated rate of false positives in the literature \citep{agnoli}, and the lack of a systematic approach for the description of methods, computer code, and data analysis in publications across all fields \citep{nuijten}.  An earlier essay in \textit{PLOS Medicine} carried the provocative title, ``Why Most Published Research Findings Are False'' \citep{Ioannidis}.  In addition, in 2013, a cover story in the popular magazine, \textit{The Economist}, invited readers to learn ``How Science Goes Wrong'' and Richard Harris’s popular 2017 book \textit{Rigor Mortis} provided many examples of purported failures in science. 

While we might be inclined to use the terms ``reproduce'' and ``replicate'' interchangeably, these two terms are in fact distinct.  Reproducibility is the ability to reproduce the results of another researcher beginning with their computer code and data, while replicability is the ability of independent researchers to collect new study data and verify the results (different groups of researchers are using different terminologies sometimes in utter contradiction with each other, but for more details on terminology see \citealp{barba}).  The objective of replicability is to quickly repudiate spurious results and enforce a ruled based approach to scientific discovery.  Both terms facilitate the ongoing self-correcting nature of science.  Indeed, a new scientific discovery requires both confirmation and extensive retesting in order to study the limits of the original result.  

While replicability is generally regarded as the gold standard of verifying the result from a scientific study, it is often difficult to perform for many reasons including experimental costs, experimental length, recreating experimental conditions, inherent variability in the system, inability to control complex variables, and substandard research practices. Hence, reproducibility is the compromise.    In fact, reproducibility can be used interchangeably with computational reproducibility \citep{stodden2018empirical}, which is embedded in numerous disciplines due to the ever growing capabilities of modern computation.  As the development of computational algorithms increases, the research community assumes the computational component of the work can be easily reproduced by not only the original authors but by other researchers too.  However, the reality is very different.  There are a multitude of problems in reproducible research.   One common problem, for example, is the lack of detailed instructions for how the analyses (the computer code with the accompanying data) should be performed.  Many of the workflows that are used to derive the results are highly customized which, in combination with the often-limited information provided in the corresponding paper, make analyses hard to reproduce.  Also, many computational algorithms remain opaque due to their increasing complexity.   This makes the documentation and hence the reproduction both cumbersome and difficult.   The level of detail required to reproduce the computational analysis is often not reported in the published paper, or the analysis is immensely time consuming.  Additionally, the final computer code, script or data for producing the final analysis may be lost by or is unrecoverable from the authors.   Furthermore, large data sets themselves, due to their size, are infeasible to process without access to specialized computer resources \citep{BOULUND201881}. The most frequently occurring issues associated with reproducibility can be summarized into four main points: (1) access to the real data; (2) availability of the final version of executable computation codes; (3) full details of the analysis workflow, and (4) complete description of the computer environment (and information on software versions) that was used to calculate the results.  Consequently, many researchers have concluded that a credibility crisis is occurring in the field of computational analysis \citep{donoho}. 

The field of statistics prides itself on its openness when it comes to sharing both computer code and data.  In addition, anecdotally, in our experience, statisticians are very responsive to sharing computer code and data when requested by email.  However, there is no quality control in the sharing of code and data.  While there is currently a great many research papers on reproducibility in other computational fields, there has been no study, to the best of our knowledge, on the reproducibility of results in statistics.  To this end, in this paper, we examine the current status of reproducibility in statistics by attempting to reproduce the results in seven prominent journals:  the \textit{Annals of Applied Statistics}, \textit{Biometrics}, \textit{Biostatistics},  the \textit{Journal of Computational and Graphical Statistics}, the \textit{Journal of the American Statistical Association}, the \textit{Journal of the Royal Statistical Society: Series C}, and \textit{Statistics in Medicine} during the period 2010-2021.  Many of these journals are currently in the process of revising author guidelines to include computer code and data availability.  In the language of \cite{stodden2015reproducing}, we are focusing on computational reproducibility, which refers to ``changes in scientific practice and reporting standards to accommodate the use of computational technology occurring primarily over the past two decades, in particular whether the same results can be obtained from the data and code used in the original study''.  Each journal publishes clear ``Requirement for codes'' and ``Requirement for data'' instructions for authors on their websites and ``encourage'' or ``strongly encourage'' authors to provide the paper-related materials including computation sources and data sets.  Some of the journals also provide data archiving services for the convenience of data upload and management.  Badges in recognition of outstanding contributions to open research have been established as well. However, such attempts have not received equal returns so far.

We focus on all published papers utilizing functional magnetic resonance imaging (fMRI) data from the journals during the period (93 papers in total).  fMRI is a valuable tool for studying neural activity in the central nervous system due to its wide spatial coverage and non-invasive nature.  Essentially, each fMRI data set has 4 dimensions, including spatial and temporal information of the Blood Oxygenation Level Dependency (BOLD) signal, which measures the neural activity by reflecting changes in blood flow.   At first glance, the results from statistical methods applied to fMRI studies are expected to be reproducible as long as the computer code and preprocessed fMRI data are provided. However, we find that statistical papers on fMRI data in the recent 11 years are often  not reproducible.  In fact, among all the 93 examined papers, we could only reproduce the results in 14 (15.1\%) papers, that is, the papers provide both executable computer code (or software) with the real fMRI data, and our results matched the results in the paper.  The failure to reproduce results is often due to i) incomplete, outdated, or missing instructions for running the computer code or software; ii) missing, outdated, inexecutable or unannotated source code files;  and/or iii) missing fMRI data or raw fMRI data that had not been preprocessed (or the failure to provide the preprocessing script).  Without well-annotated computer source code, it is very difficult for researchers to reproduce the result from scratch. Therefore it relies fully on the descriptions provided in the publications, which are often incomplete and prone to errors.  Furthermore, many authors prefer to provide access to raw, publicly-available fMRI data sets rather than directly present the preprocessed fMRI data or offer the raw data with their preprocessing code or preprocessing pipeline.  Since no agreement has been reached in terms of the best-preprocessed pipeline for fMRI data, ambiguously dealing with the raw fMRI data also jeopardizes the reproducibility of the paper. 

The remainder of this paper is organized as follows.  We summarize the reproducibility results of the 93 published papers based on fMRI data in 7 prominent statistics journals and relate these results to the computer code and data requirements from the corresponding journal in Section 2.  We discuss the availability of computer code and data for each specific journal in Section 3.  Finally, in Section 4, we detail some author-specific and journal-specific suggestions to improve research reproducibility in statistics and in computational methods in general and discuss the strengths and limitations of our own study.  We also discuss some of the many initiatives the journals have proposed (and in many cases implemented) to improve reproducibility in statistics. 

%=================================================================================================
\section{Related work}
Reproducibility and replicability have been studied in other fields including in political science \citep{king}, econometric research \citep{koenker},  operations research \citep{nestler}, archaeology \citep{marwick}, chemical engineering \citep{han}, economics \citep{vilhuber}, transportation \citep{zheng}, evolutionary computation \citep{lopez}, physics \citep{clementi}, and computational biology~\citep{cadwallader}.

While there is currently a great deal of research papers on reproducibility in other computational fields, there has been no study, to the best of our knowledge, on the reproducibility of results in the field of statistics (\citealt{stodden2018empirical} studied computational reproducibility for papers published in the journal, \textit{Science}).   There are, however, some related papers.  For example, \cite{gentleman} described a software framework for both authoring and distributing integrated, dynamic documents that contain text, code, data, and any auxiliary content needed to recreate the computations.   In an editorial for the journal \textit{Biostatistics},  \cite{peng2009reproducible} described the difficulties in and the efforts to promote reproducibility in biostatistical research.  \cite{deangelis} discussed the importance of independent statistical analysis of industry-sponsored studies, a requirement for the \textit{Journal of the American Medical Association}.  \cite{schulte} introduced a multi-language computing environment for literate programming and reproducible research.   The phyloseq project \citep{mcmurdie} is an open-source R software tool for statistical analysis of phylogenetic sequencing data, which enables reproducible preprocessing, analysis, and publication-quality graphics production.   \cite{xie} created the R package \textbf{knitr}, which combines computer code and software documentation in the same document that allows for easier reproducibility.  \cite{stodden2015reproducing} provided an overview of issues of reproducibility and how statistical research has been and could be addressing these concerns.   In \cite{fuentes}, the editor of the \textit{Journal of the American Statistical Association}, Applications and Case Studies, introduced the reproducibility initiative as a response of the reproducibility/replication crisis in science.  The author noted that most statistical papers did not submit adequate supporting computer code or data that enabled reproduction of their results.  \cite{leek} described the range of definitions of false discoveries in the scientific literature and summarize the philosophical, statistical, and experimental evidence for each type of false discovery.  To address the challenge of reproducibility with increasing computer complexity, \cite{marwick2018} reviewed the concept of the research compendium as a solution for providing a standard and easily recognizable way for organizing the digital materials of a research project to enable other researchers to inspect, reproduce, and extend the research.  \cite{becker} presented the \textbf{trackr} and \textbf{histry} R packages.  Together, these packages define a framework for tracking, automatically annotating, discovering, and reproducing the intermediate and final results of computational work done within R.   In \cite{benjamini}, the authors argued that addressing selective inference is a missing statistical cornerstone of enhancing replicability.   Related to this work, \cite{hung} applied multiple testing and post selection inference techniques to develop new
statistical methods for replicability assessment.  To increase the implementation of reproducible research in data science projects in R and to provide standards on reproducibility in published research, \cite{bertin} presented the R package \textbf{fertile}, that proactively prevents reproducibility mistakes from happening in the first place, and retroactively analyzes code for potential problems.

%===========================================================================================================
\section{Reproducibility in statistics journals}
In this paper, we explore the reproducibility of the results from applied and methodological statistical papers based on functional magnetic resonance imaging (fMRI) data published in seven prominent statistical journals during the 2010-2021 time period.   In total, we identified 93 eligible papers (obviously, this is continuously subject to change but we intend to add new papers as they are published).  We first identified the journals, and then we inspected each issue from 2010 to 2021 for fMRI data.   Multiple human readers were used to confirm the availability of code and data.  All the journals provide detailed descriptions of the computer code and data requirements, which we  summarize in Table \ref{requirements}.  We also include website links for each journal, where the requirements are provided.  
Most journals use the words `encourage', `expect', `should' in their requirements with respect to computer code and data, but in most cases they are not required. 

\footnotesize
\begin{table}[htbp]
\footnotesize{
    \begin{tabular}{p{4.5em}|p{18em}|l|p{4.5em}}
    \toprule
    Journal & \multicolumn{1}{l|}{Requirement for data} & Requirement for codes & \multicolumn{1}{l}{URL} \\
    \midrule
        AOAS & AOAS \textbf{strongly encourages} authors to make the data used in papers published in AOAS available for others to analyze. & \multicolumn{1}{p{21.665em}|}{Authors are \textbf{encouraged} to utilize web-based supplementary files to include software, or code for carrying out the analyses presented in a paper. } &
        Click \href{        https://imstat.org/journals-and-publications/annals-of-applied-statistics/annals-of-applied-statistics-manuscript-submission/}{\textcolor{blue}{here}} \\
    \midrule
    Biometrics & Biometrics \textbf{encourages} authors to share the data supporting the results in their study by archiving them in an appropriate public repository. Biometrics also \textbf{encourages} authors to submit data used in their illustrative examples if at all possible (along with code used for the analysis). & \multicolumn{1}{p{21.665em}|}{Biometrics \textbf{strongly encourages} authors to include software implementing proposed methodology with their papers at the time of submission, such as code implementing simulations or data analyses presented in the paper or, preferably, more generic software (e.g., a R package or SAS macro).} &
    Click \href{https://biometrics.biometricsociety.org/home/author-guidelines#h.p_GNcAvEniYxGa}{\textcolor{blue}{here}} \\
    \midrule
    Biostatistics & There is the opportunity to present extensive analyses of data on the journal's website as supplementary material. & \multicolumn{1}{p{21.665em}|}{Authors are \textbf{strongly encouraged} to submit code supporting their publications. Authors should submit a link to a Github repository and to a specific example of the code on a code archiving service such as Figshare or Zenodo.} &
    Click \href{https://academic.oup.com/biostatistics/pages/About}{\textcolor{blue}{here}} \\
    \midrule
    JCGS & \multicolumn{2}{p{39.665em}|}{Authors are \textbf{expected} to submit code and datasets as online supplements to the manuscript. Exceptions for reasons of security or confidentiality may be granted by the Editor. } &
    Click \href{https://amstat.tandfonline.com/action/authorSubmission?show=instructions&journalCode=ucgs20}{\textcolor{blue}{here}} \\
    
      \midrule
  \multirow{2}[4]{*}{JASA} & \multicolumn{2}{p{39.665em}|}{\textcolor{red}{Before September 1, 2021}: The ASA \textbf{strongly encourages} all authors to submit datasets, code,other programs, and/or extended appendices that are directly relevant to their submitted articles to Theory \& Methods. Since \textcolor{red}{September 1, 2016} authors publishing in the Applications and Case Studies section of JASA will be \textbf{asked} to provide materials that demonstrate reproducibility.}  & Click \href{https://www.tandfonline.com/action/authorSubmission?show=instructions&journalCode=uasa20#style}{\textcolor{blue}{here}} \\
\cmidrule{2-4}      & \multicolumn{2}{p{39.665em}|}{\textcolor{red}{After September 1, 2021}: \textbf{All} invited revisions to JASA (both Applications \& Case Studies and Theory \& Methods) for manuscripts whose initial submission was on or after September 1, 2021, \textbf{must} include code, data, and the workflow to reproduce the work presented. Published papers will include a link to reviewed reproducibility materials, including the Author Contributions Checklist; the materials will be posted to the JASA GitHub repository.} & Click \href{https://jasa-acs.github.io/repro-guide/}{\textcolor{blue}{here}} \\

    \midrule
    JRSS,C & \multicolumn{2}{p{39.665em}|}{It is the policy of the Journal of the Royal Statistical Society that published papers \textbf{should}, where possible, be accompanied by the data and computer code used in the analysis. Both data and code must be clearly and precisely documented, in enough detail that it is possible to replicate all results in the final version of the paper.} &
    Click \href{https://rss.onlinelibrary.wiley.com/hub/journal/14679876/author-guidelines}{\textcolor{blue}{here}} 
    \\
    \midrule
    Statistics in Medicine & Statistics in Medicine \textbf{expects} that data supporting the results reported in the paper will be archived in an appropriate public repository.  & \multicolumn{1}{p{21.665em}|}{The journal \textbf{requires} authors to supply any supporting computer code or simulations that allow readers to institute any new methodology proposed in the published article.} & 
    Click \href{https://onlinelibrary.wiley.com/page/journal/10970258/homepage/forauthors.html}{\textcolor{blue}{here}} \\
    
    \bottomrule
    \end{tabular}
    }
      \caption{Computer code and data requirements as stated on the websites of the seven statistical journals:  the \textit{Annals of Applied Statistics}, 
\textit{Biometrics}, \textit{Biostatistics},  the \textit{Journal of Computational and Graphical Statistics}, the \textit{Journal of the American Statistical Association}, the \textit{Journal of the Royal Statistical Society: Series C} and \textit{Statistics in Medicine}.  A url link for each journal's requirements is also provided. }
  \label{requirements}
\end{table}

\normalsize 
Table \ref{result} summarizes the results of the reproducibility of the papers from the computer code and the data perspective.   We took a lenient definition of reproducibility due to the possibility of randomness in the statistical algorithms.  For computer code, we checked whether the paper provided scripts, a package (with and without a paper script), or no computer code.  We considered whether the provided code worked, failed due to errors in the code, or failed due to an executable error (files missing in the script or whether the software package did not not include a script). For data, we checked whether the paper provided preprocessed fMRI data, simulated data, raw data (with and without a script for the preprocessing), or no data. 
\begin{table}[htbp]
  \centering
    \begin{tabular}{|p{4.045em}c|ccc|c|c|}
    \toprule
    \multicolumn{2}{|c|}{\multirow{2}[4]{*}{}} & \multicolumn{3}{c|}{code provided} & \multicolumn{1}{c|}{\multirow{2}[4]{*}{code not provided}} & \multicolumn{1}{c|}{\multirow{2}[4]{*}{total}} \\
\cmidrule{3-5}    \multicolumn{2}{|c|}{} & \multicolumn{1}{c|}{code working} & \multicolumn{1}{c|}{code failed} & \multicolumn{1}{c|}{code not executable} &   &  \\
    \midrule
    \multicolumn{1}{|c|}{\multirow{3}[6]{*}{data\newline{} provided}} & \multicolumn{1}{c|}{real data } & \multicolumn{1}{c|}{14(1)} & \multicolumn{1}{c|}{2} & \multicolumn{1}{c|}{4(4)} & 3 & \multirow{3}[6]{*}{\textbf{51}} \\
\cmidrule{2-6}    \multicolumn{1}{|c|}{} & \multicolumn{1}{c|}{sim data} & \multicolumn{1}{c|}{12(2)} & \multicolumn{1}{c|}{3} & 1 & 0 &  \\
\cmidrule{2-6}    \multicolumn{1}{|c|}{} & \multicolumn{1}{c|}{raw data} & \multicolumn{1}{c|}{0} & \multicolumn{1}{c|}{0} & \multicolumn{1}{c|}{4(2)} & 8 &  \\
    \midrule
    \multicolumn{2}{|c|}{data not provided} & \multicolumn{1}{c|}{0} & \multicolumn{1}{c|}{0} & \multicolumn{1}{c|}{7(1)} & 35 & \textbf{42} \\
    \midrule
    \multicolumn{2}{|c|}{total} & \multicolumn{3}{c|}{\textbf{47}} & \textbf{46} & \textbf{93} \\
    \bottomrule
    \end{tabular}%

      \vspace{0.5cm}
      
  \caption{The reproducibility results from a computer code and a data perspective from applied and methodological statistical papers based on functional magnetic resonance imaging (fMRI) data published in seven prominent statistical journals: the \textit{Annals of Applied Statistics}, \textit{Biometrics}, \textit{Biostatistics},  the \textit{Journal of Computational and Graphical Statistics}, the \textit{Journal of the American Statistical Association}, the \textit{Journal of the Royal Statistical Society: Series C} and \textit{Statistics in Medicine} from 2010 to 2021. The numbers in the parenthesis represent the number of R software packages in that cell.}
  \label{result}%
\end{table}%

Overall, we find that 46 out of the 93 (49\%) published papers provide no computer code (or software).  Out of the 47 published papers with computer code available, we find that 26 (28\%) provide computer code that runs smoothly, 5 (5\%) where the code provided failed, and 16 (17\%) where the code included was not executable (e.g., only the R functions were provided but there is missing key files).  In addition, 10 papers provide an R software package \citep{R}, of which only 3 contain user-friendly scripts to reproduce the results in the paper.  R packages are very convenient as they allow for the easy adaption of the code for application to other data sets and for easy comparison with other methods, however, ideally a paper based script should be included for reproducibility. For all cases where the computer code failed, we endeavored to fix all the errors (both major and minor). As long as the code could be executed after fixation, we classified it as `code working'.  However, most computer code that generated an error was due to missing data or missing functions, which we were not able to resolve without the help of the original authors.  From the data perspective, we find that 42 out of the 93 (45\%) published papers provide no real, simulated, or raw data.  Out of the 51 (55\%) published papers with data available, we find that 23 (25\%) provide the real fMRI data analyzed in the paper, 16 (17\%) provide simulated data, and 12 (13\%) provide the raw fMRI data.  If papers included both simulated and real data, we classified them under the real data category so we did not double count them.  

From both the computer code and data viewpoint, among the 10 papers that developed a R software package, only 1 includes the preprocessed fMRI data set in the package with clear executable instructions.  Out of the 11 papers that share the data (real, simulated or raw) but do not include any relevant computer code, 8 of them provide a link to a public website containing the raw fMRI data set.  While raw fMRI data is preferable to providing no data, it puts the onus on the researcher attempting to reproduce the results to preprocess the fMRI data.  As we discuss later, for fMRI data, it is extremely difficult to obtain precisely the same preprocessed data from the raw version as there is not an established framework of carrying out the preprocessing steps.  Even worse, several researchers do not provide full preprocessing steps in their papers. Of course, it would be acceptable if the authors provided both the raw data and a clear step-by-step script for preprocessing but even then this makes reproducing the work more difficult due to this extra step and the possibility of errors.  There are only 2 papers that provide a raw fMRI data set and a preprocessing script for the data.  However, we were unable to preprocess the raw data; therefore they are listed under `code not executable'.  Hence, from both the computer code and the data perspective, among all the 93 examined papers, only 14 papers provide executable computer code (or software) and real fMRI data, where we were able to reproduce the results. This equates to just around 15\% of the papers examined.  Table \ref{resultofcode} (in the Appendix) provides more specific details on the reproducibility of the 47 papers that provide computer code (or software). From a computer software viewpoint, R is the mostly used software (33/47 of the papers employ it solely or employ it in combination with another software), with Matlab second (14/47 employ it solely or employ it in combination with another software).  

\begin{figure}[p]
\centering
\includegraphics[scale = 0.9]{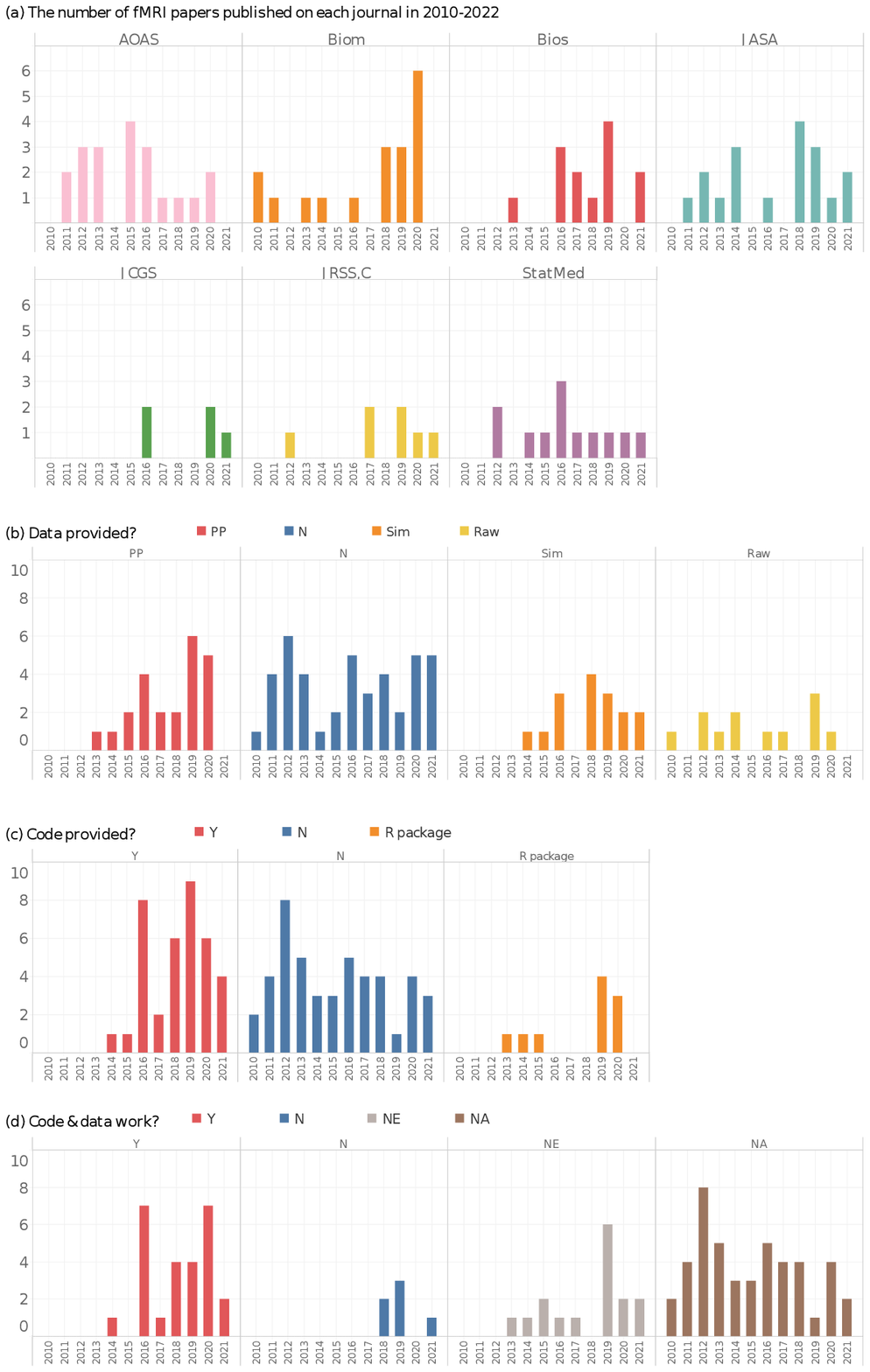}
\caption{(a) The number of fMRI papers published in the seven journals; (b) the number of papers providing preprocessed fMRI data (``PP''), no data (``N''), simulated data (``Sim''), and raw fMRI data (``Raw''); (c) the number of papers providing computer code (``Y''), no computer code (``N''), and an R software package (``R package''); (d) the number of papers providing computer code and data that can be executed, without failure (``Y''), with code errors (``N''), with code that is not executable (``NE''), or with errors due to missing data reasons (``NA''), for each year during the time period 2010-2021.}
\label{year}
\end{figure}

In Figure \ref{year}, we illustrate the timeline of the number of fMRI papers published in each journal, and their efforts towards reproducibility for each year during the period 2010-2021. From Figure \ref{year}(a), we conclude that the \textit{Annals of Applied Statistics} has published the most papers on fMRI data, while \textit{Biometrics}, \textit{Biostatistics} and \textit{JASA} appear to be publishing the most papers related to fMRI data in the recent five years.  On the contrary, \textit{AOAS} has focused less on fMRI topics recently given the decreasing number of publications.  In Figure \ref{year}(b), which displays the number of papers that provide the real preprocessed fMRI data, no data, simulated data, and raw fMRI data, we notice that more recently published papers choose to submit the ideal case, a preprocessed fMRI data set. In 2019, 86\% (12 out of 14) of the fMRI papers published in the seven journals offered at least one type of fMRI (simulated, raw, or preprocessed) data.  This positive trend does not continue, there is a slight downward trend in 2020 and 2021.  The increasing emphasis on the availability of computer code or software in recent years is also evident in Figure \ref{year}(c).  Figure \ref{year}(d) provides more information on reproducibility: it considers both computer code and data availability and  jointly reproducibility.  In our papers considered, there are many examples where both the code can be executed on the data without errors, but there are more papers where the data is not available. Although half of the fMRI papers in the journals began to archive executable computer code in combination with real or simulated data since 2018, some of the computer code files are not executable due to unexpected errors or technical problems. In fact, among the 44 papers that were published after 2018 that provided computer code, we were only able to smoothly generate outputs in 15 of them (34\%). Nevertheless, the trajectory in this plot is increasing.

In the next section, we discuss the availability of computer code and data for each specific journal.

%================================================================================
\section{Specific journals}

The authors are happy to share the table for all papers considered, with links to computer code and data, in the study but as mentioned earlier our objective is not to single out individual researchers. 

\subsection{Annals of Applied Statistics (AOAS)}
In AOAS, we identified 20 published papers related to fMRI data during the time period 2010--2021.  Out of the 20 papers, only 6 (30\%) provided executable computer code (Table~\ref{result_aoas}).  Out of these 6, 3 created an R software package and included the preprocessed fMRI data as an illustrative example therein.  We could not reproduce their results as the authors did not provide a specific script in their packages for the real data analysis.  Nevertheless, using the instruction file in the reference manual of the package, other researchers may be able to reproduce the analysis on the real fMRI data set, but it would require extensive work and interpretation.  

%AOAS
\begin{table}[htbp]
  \centering
    \begin{tabular}{|cc|ccc|c|c|}
    \toprule
    \multicolumn{2}{|c|}{\multirow{2}[4]{*}{}} & \multicolumn{3}{c|}{code provided} & \multicolumn{1}{c|}{\multirow{2}[4]{*}{code not provided}} & \multicolumn{1}{c|}{\multirow{2}[4]{*}{total}} \\
\cmidrule{3-5}    \multicolumn{2}{|c|}{} & \multicolumn{1}{c|}{code working} & \multicolumn{1}{c|}{code failed} & \multicolumn{1}{c|}{code not executable} &   &  \\
    \midrule
    \multicolumn{1}{|c|}{\multirow{3}[6]{*}{data provided}} & \multicolumn{1}{c|}{real data } & \multicolumn{1}{c|}{2} & \multicolumn{1}{c|}{} & \multicolumn{1}{c|}{3(3)} & 2 & \multirow{3}[6]{*}{\textbf{11}} \\
\cmidrule{2-6}    \multicolumn{1}{|c|}{} & \multicolumn{1}{c|}{sim data} & \multicolumn{1}{c|}{} & \multicolumn{1}{c|}{} & 1 &   &  \\
\cmidrule{2-6}    \multicolumn{1}{|c|}{} & \multicolumn{1}{c|}{raw data} & \multicolumn{1}{c|}{} & \multicolumn{1}{c|}{} &   & 3 &  \\
    \midrule
    \multicolumn{2}{|c|}{data not provided} & \multicolumn{1}{c|}{} & \multicolumn{1}{c|}{} &   & 9 & \textbf{9} \\
    \midrule
    \multicolumn{2}{|c|}{total} & \multicolumn{3}{c|}{\textbf{6}} & \textbf{14} & \textbf{20} \\
    \bottomrule
    \end{tabular}%
    
          \vspace{0.5cm}
      
  \caption{The reproducibility results from the computer code and the data perspective from applied and methodological statistical papers based on functional magnetic resonance imaging (fMRI) data published in the \textit{Annals of Applied Statistics} from 2010 to 2021. The numbers in the parenthesis represent the number of R software packages in that cell.}
\label{result_aoas}
\end{table}%

From the data perspective, out of the 20 published papers, 7 (35\%) attached the preprocessed data (or at least some of the preprocessed data in the supplementary materials), 1 provided simulated data, 3 mentioned the website where the raw data can be downloaded (but did not provide a preprocessing script), and the remaining 9 did not mention data availability. The high missing proportion (9/20) may be attributed to the early publication date of these papers: 5 were published before 2014 when the topic of reproducibility was not as important an issue generating a significant amount of coverage and discussion.  Among the more recently published papers, there are only 2 papers that provide both computer code and the preprocessed fMRI data sets, but we were unable to reproduce precisely the same result as detailed in the published paper in both of them. One paper missed a key brain file used in the computer code, and in the other, we detected a different number of significant region of interest-single nucleotide polymorphisms (ROI-SNP) connections (however, we still defined this as reproducible due to our lenient interpretation). 

Hence, overall, from both the computer code and the data perspective, among all the 20 examined papers in AOAS, only 2 (10\%) paper provides executable computer code (or software) and real fMRI data, where we were able to reproduce the results.

\subsection{Biometrics}
In \textit{Biometrics}, we identified 18 published papers related to fMRI data during the time period 2010--2021.  Out of the 18 papers, 14 (78\%) shared computer code (Table~\ref{result_biometrics}).  Out of those 14, 5 papers provided software packages implementing the proposed methodology.  \textit{Biometrics} is the only journal among the seven studied that states their preference for generic software (e.g., R packages or SAS macro) over executable computer code implementing simulations or data analyses.  However, the journal did not require authors to include the real data set used in the paper in the software, which resulted in 1 package having no illustrative data example, 2 packages with only simulated data sets, and the remaining 2 packages with preprocessed fMRI data.  For the 9 published papers that provide computer code (as scripts), 1 provided R functions without executable lines, 1 failed to run due to a missing file, and the remaining 7 produced reasonable results.  
% Biometrics
\begin{table}[htbp]
  \centering
    \begin{tabular}{|cc|ccc|c|c|}
    \toprule
    \multicolumn{2}{|c|}{\multirow{2}[4]{*}{}} & \multicolumn{3}{c|}{code provided} & \multicolumn{1}{c|}{\multirow{2}[4]{*}{code not provided}} & \multicolumn{1}{c|}{\multirow{2}[4]{*}{total}} \\
\cmidrule{3-5}    \multicolumn{2}{|c|}{} & \multicolumn{1}{c|}{code working} & \multicolumn{1}{c|}{code failed} & \multicolumn{1}{c|}{code not executable} &   &  \\
    \midrule
    \multicolumn{1}{|c|}{\multirow{3}[6]{*}{data provided}} & \multicolumn{1}{c|}{real data } & \multicolumn{1}{c|}{5(1)} & \multicolumn{1}{c|}{} & \multicolumn{1}{c|}{1(1)} &   & \multirow{3}[6]{*}{\textbf{13}} \\
\cmidrule{2-6}    \multicolumn{1}{|c|}{} & \multicolumn{1}{c|}{sim data} & \multicolumn{1}{c|}{5(2)} & \multicolumn{1}{c|}{1} &   &   &  \\
\cmidrule{2-6}    \multicolumn{1}{|c|}{} & \multicolumn{1}{c|}{raw data} & \multicolumn{1}{c|}{} & \multicolumn{1}{c|}{} &   & 1 &  \\
    \midrule
    \multicolumn{2}{|c|}{data not provided} & \multicolumn{1}{c|}{} & \multicolumn{1}{c|}{} & \multicolumn{1}{c|}{2(1)} & 3 & \textbf{5} \\
    \midrule
    \multicolumn{2}{|c|}{total} & \multicolumn{3}{c|}{\textbf{14}} & \textbf{4} & \textbf{18} \\
    \bottomrule
    \end{tabular}%
              
              \vspace{0.5cm}
      
  \caption{The reproducibility results from the computer code and the data perspective from applied and methodological statistical papers based on functional magnetic resonance imaging (fMRI) data published in \textit{Biometrics} from 2010 to 2021. The numbers in the parenthesis represent the number of R software packages in that cell.}
\label{result_biometrics}
\end{table}%

From the data perspective, 13 (72\%) of the 18 papers provided some data related to the paper.  In particular, 6 provided preprocessed data, 6 chose to share simulated data, 1 offered links to raw data (with no preprocessing script).  The remaining 5 papers provided no information on the data source.  Overall, published papers in \textit{Biometrics} provide very good capacity for reproducibility: from both the computer code and the data perspective, among all the 18 examined papers, 5 (28\%) papers provide executable computer code (or software) and real preprocessed fMRI data, which we were able to reproduce the results.  The 3 papers missing both the computer code and data resources were published before 2014. One specific paper that was published recently claims that the code is available on the \textit{Biometrics} website, however, it only provides a link to the raw fMRI data without the computer code for preprocessing. 

\subsection{Biostatistics}
In \textit{Biostatistics}, we identified 13 published papers related to fMRI data during the time period 2010--2021.    The majority (10/13) of these papers were published after 2015. From the computer code perspective, 8 (62\%) papers attached computer code, among which 1 created an R package (Table~\ref{result_bios}).  However, only 4 of the 8 computer codes were in working order.   The paper with an R package did include some sample data sets in the software, however, no specific paper-related scripts that would allow reproduction is provided.

% Biostatistics
\begin{table}[htbp]
  \centering
    \begin{tabular}{|cc|ccc|c|c|}
    \toprule
    \multicolumn{2}{|c|}{\multirow{2}[4]{*}{}} & \multicolumn{3}{c|}{code provided} & \multicolumn{1}{c|}{\multirow{2}[4]{*}{code not provided}} & \multicolumn{1}{c|}{\multirow{2}[4]{*}{total}} \\
\cmidrule{3-5}    \multicolumn{2}{|c|}{} & \multicolumn{1}{c|}{code working} & \multicolumn{1}{c|}{code failed} & \multicolumn{1}{c|}{code not executable} &   &  \\
    \midrule
    \multicolumn{1}{|c|}{\multirow{3}[6]{*}{data provided}} & \multicolumn{1}{c|}{real data } & \multicolumn{1}{c|}{2} & \multicolumn{1}{c|}{1} &   &   & \multirow{3}[6]{*}{\textbf{10}} \\
\cmidrule{2-6}    \multicolumn{1}{|c|}{} & \multicolumn{1}{c|}{sim data} & \multicolumn{1}{c|}{2} & \multicolumn{1}{c|}{1} &   &   &  \\
\cmidrule{2-6}    \multicolumn{1}{|c|}{} & \multicolumn{1}{c|}{raw data} & \multicolumn{1}{c|}{} & \multicolumn{1}{c|}{} & \multicolumn{1}{c|}{1(1)} & 3 &  \\
    \midrule
    \multicolumn{2}{|c|}{data not provided} & \multicolumn{1}{c|}{} & \multicolumn{1}{c|}{} & 1 & 2 & \textbf{3} \\
    \midrule
    \multicolumn{2}{|c|}{total} & \multicolumn{3}{c|}{\textbf{8}} & \textbf{5} & \textbf{13} \\
    \bottomrule
    \end{tabular}%

              \vspace{0.5cm}
      
  \caption{The reproducibility results from the computer code and the data perspective from applied and methodological statistical papers based on functional magnetic resonance imaging (fMRI) data published in \textit{Biostatistics} from 2010 to 2021. The numbers in the parenthesis represent the number of R software packages in that cell.}
\label{result_bios}
\end{table}%

As for data, 10 (77\%) of the published papers are accompanied by some form of fMRI data used in the analysis, 4 of which only provide a link to a website with open source raw fMRI data (one paper pointed to its R package for its preprocessing script, but we could not execute it). As mentioned before, a unique trait of fMRI data is that is has to be preprocessed, but no established sequence in carrying out the preprocessing steps exists.  Hence, while raw fMRI data is preferable to no data, it puts the onus on the researcher attempting to reproduce the results to preprocess the fMRI data, which is extremely difficult.  Even if researchers describe their preprocessing pipeline in great detail, matching the final data in the published paper cannot be guaranteed.  Without providing a reason, 3 published papers preferred to provide simulated data rather than the complete fMRI data analyzed in the paper.  This may be due to proprietary nature of fMRI data.  The simulated version of the data and the corresponding computer code may be due to the requirements of \textit{Biostatistics} (see Table \ref{requirements}), which states its encouragement to submit computer code for a specific example, rather than the exact real data set.  Nevertheless, for these papers we could at least generate readable results, compared to another published paper which attached the computer code that was rigidly designed for the real data analysis but failed to provide the real data.  Furthermore, 2 published papers refer to an fMRI study of thermal pain but neither of them provided the data source.   

Overall, from both the computer code and the data perspective, among all the 13 examined papers in \textit{Biostatistics}, only 2 (15\%) papers provided properly organized computer code and real fMRI data, with which we were able to reproduce the results. Particularly, one paper partitions the fMRI time series ($T=197$) into two segments and only shares a truncated portion ($t \in [0,50]$) of it.  \textit{Biostatistics} has been concerned with reproducibility since 2009 when its Editors announced its computational reproducibility policy \citep{peng2009reproducible} to promote reproducibility in biostatistical research.  Here, after an article has been accepted for publication, the assigned Associate Editor for reproducibility (AER),  considers three different criteria (Data, Code, Reproducible) when evaluating the reproducibility of an article.  Published papers that meet any or all of the three criteria are marked D, C, and/or R on their title page in the journal.  However, this process is not mandatory, only optional, and does not extent beyond the R software. 

\subsection{Journal of Computational and Graphical Statistics (JCGS)}
In JCGS, we identified only 5 published papers related to fMRI data during the time period 2010--2021.  Most of the papers provide detailed documents explaining the implementation of the computer code in the supplementary materials except one that published recently in 2021.  In particular, 3 of the 5 (60\%) papers provide working computer code, 1 provides code that is not executable, while the final paper provides no code (Table~\ref{result_jcgs}).  

% JCGS
\begin{table}[htbp]
  \centering
    \begin{tabular}{|cc|ccc|c|c|}
    \toprule
    \multicolumn{2}{|c|}{\multirow{2}[4]{*}{}} & \multicolumn{3}{c|}{code provided} & \multicolumn{1}{c|}{\multirow{2}[4]{*}{code not provided}} & \multicolumn{1}{c|}{\multirow{2}[4]{*}{total}} \\
\cmidrule{3-5}    \multicolumn{2}{|c|}{} & \multicolumn{1}{c|}{code working} & \multicolumn{1}{c|}{code failed} & \multicolumn{1}{c|}{code not executable} &   &  \\
    \midrule
    \multicolumn{1}{|c|}{\multirow{3}[6]{*}{data provided}} & \multicolumn{1}{c|}{real data } & \multicolumn{1}{c|}{2} & \multicolumn{1}{c|}{} &   &   & \multirow{3}[6]{*}{\textbf{3}} \\
\cmidrule{2-6}    \multicolumn{1}{|c|}{} & \multicolumn{1}{c|}{sim data} & \multicolumn{1}{c|}{1} & \multicolumn{1}{c|}{} &   &   &  \\
\cmidrule{2-6}    \multicolumn{1}{|c|}{} & \multicolumn{1}{c|}{raw data} & \multicolumn{1}{c|}{} & \multicolumn{1}{c|}{} &   &   &  \\
    \midrule
    \multicolumn{2}{|c|}{data not provided} & \multicolumn{1}{c|}{} & \multicolumn{1}{c|}{} & 1 & 1 & \textbf{2} \\
    \midrule
    \multicolumn{2}{|c|}{total} & \multicolumn{3}{c|}{\textbf{4}} & \textbf{1} & \textbf{5} \\
    \bottomrule
    \end{tabular}%
    
                  \vspace{0.5cm}
      
  \caption{The reproducibility results from the computer code and the data perspective from applied and methodological statistical papers based on functional magnetic resonance imaging (fMRI) data published in the \textit{Journal of Computational and Graphical Statistics} from 2010 to 2021. The numbers in the parenthesis represent the number of R software packages in that cell.}
  \label{result_jcgs}
\end{table}%

From the data perspective, 2 of the published papers attach the preprocessed fMRI data and one offers simulated data sets (60\% in total) and all can be combined with the code and executed smoothly.  Hence, overall, from both the computer code and the data perspective, among all the 5 examined papers in JCGS, 2 (40\%) papers provide properly organized computer code and real preprocessed fMRI data, where we were able to reproduce the results.  Compared to the other journals in this study, JCGS performs the best in terms of making the computer code and data available for reproduction.  However, the number of papers is small.

\subsection{Journal of the American Statistical Association (JASA)} \label{subsec:jasa}
In JASA, we identified 18 published papers related to fMRI data during the time period 2010--2021.  Out of the 18 papers, only 9 of them (50\%) made computer code or an R package (1 paper) relevant to the paper available.  Out of the 9 papers, 3 had working code, 1 had code that failed, and 5 had code that was not executable (including the R package) (Table~\ref{result_jasa}).

% JASA
\begin{table}[htbp]
  \centering
    \begin{tabular}{|cc|ccc|c|c|}
    \toprule
    \multicolumn{2}{|c|}{\multirow{2}[4]{*}{}} & \multicolumn{3}{c|}{code provided} & \multicolumn{1}{c|}{\multirow{2}[4]{*}{code not provided}} & \multicolumn{1}{c|}{\multirow{2}[4]{*}{total}} \\
\cmidrule{3-5}    \multicolumn{2}{|c|}{} & \multicolumn{1}{c|}{code working} & \multicolumn{1}{c|}{code failed} & \multicolumn{1}{c|}{code not executable} &   &  \\
    \midrule
    \multicolumn{1}{|c|}{\multirow{3}[6]{*}{data provided}} & \multicolumn{1}{c|}{real data } & \multicolumn{1}{c|}{} & \multicolumn{1}{c|}{} &   &   & \multirow{3}[6]{*}{\textbf{8}} \\
\cmidrule{2-6}    \multicolumn{1}{|c|}{} & \multicolumn{1}{c|}{sim data} & \multicolumn{1}{c|}{3} & \multicolumn{1}{c|}{1} &   &   &  \\
\cmidrule{2-6}    \multicolumn{1}{|c|}{} & \multicolumn{1}{c|}{raw data} & \multicolumn{1}{c|}{} & \multicolumn{1}{c|}{} & \multicolumn{1}{c|}{3(1)} & 1 &  \\
    \midrule
    \multicolumn{2}{|c|}{data not provided} & \multicolumn{1}{c|}{} & \multicolumn{1}{c|}{} & 2 & 8 & \textbf{10} \\
    \midrule
    \multicolumn{2}{|c|}{total} & \multicolumn{3}{c|}{\textbf{9}} & \textbf{9} & \textbf{18} \\
    \bottomrule
    \end{tabular}%
    
                  \vspace{0.5cm}
      
                  \vspace{0.5cm}
      
  \caption{The reproducibility results from the computer code and the data perspective from applied and methodological statistical papers based on functional magnetic resonance imaging (fMRI) data published in the \textit{Journal of the American Statistical Association} from 2010 to 2021. The numbers in the parenthesis represent the number of R software packages in that cell.}
  \label{result_jasa}
\end{table}%

In terms of data, 8 out of the 18 papers (44\%) provided some forms of data (real, simulated, or raw).  In particular, 4 papers chose to share simulated data to illustrate their algorithms, three of which can be reproduced without fatal errors.  Three papers provide rich materials that demonstrate in theory the reproducibility of their results and detail the source of data used to perform the analysis, as required by the journal in Table \ref{requirements}.  However, following one paper's instruction steps, we had difficulty in finding a key data file that was no longer available on the original website or in their supplementary materials. Surprisingly, none of the 18 papers directly attach the full preprocessed data, which makes reproducing real data analysis more difficult.  Although the journal stipulates a strong demand for the documentation of the source of the data as well as attaching the computer code or software, 8 of the 18 published papers provide neither in their final versions.  These results were unexpected given the prominent stature of JASA among statisticians. In fact, for the 3 papers published after January 2020, we encountered different kinds of computer code and/or data issues with all of them.  Specifically, 1 paper did not mention either the computer code or data used at all, while another only presented R functions without any other instructions or data.  Finally, while the third paper detailed the reproduction steps and provided a list of file names used to generate the figures, the zip file available for download from the website is not structured in the same manner as mentioned in the supplementary materials.  The most important Matlab toolbox implementing the method was missing, which made reproducibility impossible.  Hence, overall, from both the computer code and the data perspective, among all the 18 examined papers in JASA, no papers provide properly organized computer code and real fMRI data.

As detailed in Table~\ref{requirements}, as of September 2021, JASA (Theory and Methods, it made changes to Applications and Case Studies in 2016) has made considerable changes to its `Requirement for data' and `Requirement for codes' (link \href{https://jasa-acs.github.io/repro-guide/}{\textcolor{blue}{here}} for more details).  Specifically, the journal requires that all invited revisions ``must include code, data, and the workflow to reproduce the work presented.'' The journal also provides guidelines to authors and general resources for reproducibility.  Most significantly, the journal has stipulated that either one of the reviewers or the JASA associate editors for reproducibility (AER) will carry out a reproducibility review of the work.  Their objective is to ultimately make the reproducibility review process more efficient and rigorous (see Section~\ref{sec:conc} for more details).

\subsection{Journal of the Royal Statistical Society: Series C (JRSS,C)}
In JRSS,C, we identified 7 published papers related to fMRI data during the time period 2010--2021.  Out of the 7 papers, only 4 (57\%) provided computer code (Table ~\ref{result_jrssc}).  Out of the 4 providing computer code, 1 failed. 

% Table generated by Excel2LaTeX from sheet 'info for journals'
% JRSSc
\begin{table}[htbp]
  \centering
    \begin{tabular}{|cc|ccc|c|c|r}
\cmidrule{1-7}    \multicolumn{2}{|c|}{\multirow{2}[4]{*}{}} & \multicolumn{3}{c|}{code provided} & \multicolumn{1}{c|}{\multirow{2}[4]{*}{code not provided}} & \multicolumn{1}{c|}{\multirow{2}[4]{*}{total}} &  \\
\cmidrule{3-5}    \multicolumn{2}{|c|}{} & \multicolumn{1}{c|}{code working} & \multicolumn{1}{c|}{code failed} & \multicolumn{1}{c|}{code not executable} &   &   &  \\
\cmidrule{1-7}    \multicolumn{1}{|c|}{\multirow{3}[6]{*}{data provided}} & \multicolumn{1}{c|}{real data } & \multicolumn{1}{c|}{2} & \multicolumn{1}{c|}{1} &   &   & \multirow{3}[6]{*}{\textbf{4}} &  \\
\cmidrule{2-6}    \multicolumn{1}{|c|}{} & \multicolumn{1}{c|}{sim data} & \multicolumn{1}{c|}{1} & \multicolumn{1}{c|}{} &   &   &   &  \\
\cmidrule{2-6}    \multicolumn{1}{|c|}{} & \multicolumn{1}{c|}{raw data} & \multicolumn{1}{c|}{} & \multicolumn{1}{c|}{} &   &   &   &  \\
\cmidrule{1-7}    \multicolumn{2}{|c|}{data not provided} & \multicolumn{1}{c|}{} & \multicolumn{1}{c|}{} &   & 3 & \textbf{3} &  \\
\cmidrule{1-7}    \multicolumn{2}{|c|}{total} & \multicolumn{3}{c|}{\textbf{4}} & \textbf{3} & \textbf{7} &  \\
\cmidrule{1-7}    \end{tabular}%

                  \vspace{0.5cm}
      
  \caption{The reproducibility results from the computer code and the data perspective from applied and methodological statistical papers based on functional magnetic resonance imaging (fMRI) data published in the \textit{Journal of the Royal Statistical Society: Series C} from 2010 to 2021. The numbers in the parenthesis represent the number of R software packages in that cell.}
  \label{result_jrssc}
  \end{table}%
  
From the data perspective, 4 (57\%) out of the 7 papers provide some forms of data (real, simulated, or raw).
For the 3 papers that fail to provide any data, two of them were published recently in 2020.  Furthermore, 2 out of 7 papers provide fMRI data for analysis and executable computer code, and one of them prepares a sample of the fMRI data (one out of the 45 subjects in the original paper).  Hence, in summary, from both a computer code and data perspective, among all the 7 examined papers in JRSS,C (Applied Statistics), 2 (29\%) papers provide properly organized computer code and real preprocessed fMRI data.

Though the effort has not completely paid off yet, JRSS,C has emphasized reproducibility to a large extent.  Unlike other journals which allow authors to provide a link to computer code archiving service such as Github or to attach the computer code file in supplementary materials, JRSS,C established an open-access website
(click \href{https://rss.onlinelibrary.wiley.com/hub/journal/14679876/series-c-datasets}{\textcolor{blue}{here}}), which includes resources from its published papers dating back to 1998.  Both the computer code and the preprocessed data sets  are `clearly and precisely documented in enough detail', as required by the journal.  Nevertheless, 3 published papers in JRSS,C do not provide this for some unknown reason.

\subsection{Statistics in Medicine}
In \textit{Statistics in Medicine}, we identified 12 published papers related to fMRI data during the time period 2010--2021.  Among all the 12 published papers, only 2 (17\%) papers provide computer code (Table~\ref{result_statmed}).  In particular, only 1 paper provides executable Matlab code. The other paper attaches an R file listing all the defined functions but fails to attach a file illustrating the usage of any function. We  could not identify any computer code resource in the remaining 10 papers. 

% StatMed  
  \begin{table}[htbp]
  \centering
    \begin{tabular}{|cc|ccc|c|c|}
    \toprule
    \multicolumn{2}{|c|}{\multirow{2}[4]{*}{}} & \multicolumn{3}{c|}{code provided} & \multicolumn{1}{c|}{\multirow{2}[4]{*}{code not provided}} & \multicolumn{1}{c|}{\multirow{2}[4]{*}{total}} \\
\cmidrule{3-5}    \multicolumn{2}{|c|}{} & \multicolumn{1}{c|}{code working} & \multicolumn{1}{c|}{code failed} & \multicolumn{1}{c|}{code not executable} &   &  \\
    \midrule
    \multicolumn{1}{|c|}{\multirow{3}[6]{*}{data provided}} & \multicolumn{1}{c|}{real data} & \multicolumn{1}{c|}{1} & \multicolumn{1}{c|}{} &   & 1 & \multirow{3}[6]{*}{\textbf{2}} \\
\cmidrule{2-6}    \multicolumn{1}{|c|}{} & \multicolumn{1}{c|}{sim data} & \multicolumn{1}{c|}{} & \multicolumn{1}{c|}{} &   &   &  \\
\cmidrule{2-6}    \multicolumn{1}{|c|}{} & \multicolumn{1}{c|}{raw data} & \multicolumn{1}{c|}{} & \multicolumn{1}{c|}{} &   &   &  \\
    \midrule
    \multicolumn{2}{|c|}{data not provided} & \multicolumn{1}{c|}{} & \multicolumn{1}{c|}{} & 1 & 9 & \textbf{10} \\
    \midrule
    \multicolumn{2}{|c|}{total} & \multicolumn{3}{c|}{\textbf{2}} & \textbf{10} & \textbf{12} \\
    \bottomrule
    \end{tabular}%
                      \vspace{0.5cm}
      
  \caption{The reproducibility results from the computer code and the data perspective from applied and methodological statistical papers based on functional magnetic resonance imaging (fMRI) data published in \textit{Statistics in Medicine} from 2010 to 2021. The numbers in the parenthesis represent the number of R software packages in that cell.}
  \label{result_statmed}
  \end{table}%
  
From the data perspective, only 2 (17\%) papers provide preprocessed fMRI data sets, while 10 do not provide any data. Interestingly, out of the 12 published papers, 9 claim to provide both the computer code or the real data, but only 2 actually do so.  In one instance, a paper states in the abstract that ``software for fitting graphical object‐oriented data analysis is provided'', but the source of the software is not mentioned in the rest of the paper.  Additionally, another paper clearly states that the tool implementation is in the coding language C and adds a hyperlink, but this only links to a list of publications by the author. For a more recently published paper, the paper states at the end of the paper that `upon publication, software in the form of R code will be available from an online repository together with the sample simulated data'. This paper was first published on October 2020 but after a search online in August 2021, the data and code were still not available.

Overall, papers on the topic of fMRI published in \textit{Statistics in Medicine} do not reproduce well, with only 1 paper out of 12 (8\%) providing both computer code and the fMRI data, and we could reproduce the result. 

%==============================================================================================
\section{Conclusion}\label{sec:conc}
In this paper, we have explored the reproducibility of applied and methodological papers in the field of statistics by exploring all the papers ($n=93$) based on functional magnetic resonance imaging (fMRI) data published in seven prominent statistical journals during the time period 2010-2021.  Although statisticians pride themselves on open computer code (through the sharing of scripts or the creation of packages), we found an overall common lack of transparency and openness in both the computer code and data sets illustrating the statistical methods and applications, which raises the urgent need for attention and action.  Below, referring to the narrative in \cite{Stodden1240}, we list our recommendations for authors, editors/journals, reviewers, and funding organizations to facilitate reproducibility in statistics in general (or fMRI applications specifically) but also across other quantitative research domains.

\subsection{Author recommendations}
\noindent \textbf{Computer code}:
Instead of attaching the created functions in the supplementary materials or simply listing all the required files in one folder, we recommend that authors follow the requirements for the Application and Case Studies (ACS) section of the \textit{Journal of the American Statistical Association}.  It requests that authors provide detailed computer materials including step-by-step workflows to demonstrate reproducibility.  For example, \cite{Mejia2017} present their computer code to perform the analysis in their paper in executing orders, with a thorough explanation for the function in each step.  If possible, the visualization tool and the computation time should also be included, although these are not essential.  We also noticed that several papers share their preprocessed fMRI data (which we recommend although raw data and a precise preprocessing script is also acceptable), but their computer code fails to reproduce the result because key files are missing.  Hence, we recommend enclosing all dependent data and files (e.g., templates, brain masks, parameter settings, for fMRI data in particular) without the need to contact the authors.  In terms of the code repository, out of our 93 papers, 17 of the published papers chose GitHub, 20 submitted their codes as online supplements on the journal's webpages, while other papers archived the files on their personal websites.  It was evident that some links from the related publication on the personal websites were not accessible. We, therefore, suggest that authors share their computer code in an appropriate public repository by using persistent links (if and when they were to move to another institution or to another position).  With respect to creating software, 10 produced R software packages instead of executable R scripts, with only 3 of them attaching paper-related preprocessed fMRI data sets to the package. This makes reproduction only possible if researchers are willing to study the manual and learn how to use each R function in detail, however, this puts the onus on the (reproducing) researchers and is less convenient. Therefore we believe it is critical to provide manuals and clear paper scripts with lucid, straightforward instructions on the steps necessary to regenerate the results.

\vspace{0.5cm}

\noindent \textbf{Data}:
For all data sets, especially for fMRI data, we strongly recommend that authors provide the preprocessed data in the supplementary materials/files rather than a link to a public data website which only provides the raw version.  Specific to fMRI, as the field of neuroscience/neurostatistics has not reached an agreement on a standard preprocessing pipeline, the results can vary greatly owing to different sequences of the preprocessing steps. Although some of the papers do mention the general preprocessing steps, for example,
\begin{quote}
`We apply a series of standard image preprocessing steps: distortion-correction using FSL’s FUGUE, time-series preprocessing, rigid registration, brain extraction, temporal filtering, and 6mm FWHM Gaussian spatial smoothing. Subject-level models are fit using a linear model in FSL’s FILM software including ...' 
\end{quote}
and 
\begin{quote}
 `the preprocessing included slice time correction; 3-D motion correction; temporal despiking; spatial smoothing (FWHM=6mm); mean-based intensity normalization; temporal bandpass filtering (0.009–0.1Hz); linear and quadratic detrending ...'
\end{quote}

\noindent it is still very difficult for researchers to regenerate precisely the same data set using the identical preprocessing steps, since the parameter settings and detailed computational steps are missing.  One possible solution is to mimic the Athena strategy in the ADHD-200 Sample project
(click \href{https://www.nitrc.org/plugins/mwiki/index.php/neurobureau:AthenaPipeline#Description_of_included_files}{\textcolor{blue}{here}}).  Not only is the preprocessed data set provided, but the preprocessing script and the log file that clarify the manipulating process for each subject are also included.  Inspired by this strategy, we encourage authors that consider public fMRI data sets in their papers to, at the very minimum, provide the link to the raw data and their preprocessing script for easy access and use by other researchers.

A major issue with imaging data sets are their size, which makes the permanent storage of the data on the internet financially challenging for both the authors and journals.  However, websites such as \url{openneuro.org} offer a free and open platform for validating and sharing Brain Imaging Data Structure (BIDS)-compliant magnetic resonance imaging (MRI), positron emission tomography (PET), magnetoencephalography (MEG), electroencephalography (EEG), and intracranial Electroencephalography (iEEG) data.  The BIDS is an emerging standard for the organization of neuroimaging data, which would allow for reproducibility across research labs.  For the papers we considered in our study, when the data sets were large for local drives, the authors either parsed the data or provided a toy example.  While the ideal would be the sharing of all data and scripts, sharing a portion of the data is also acceptable.

\subsection{Suggestion for journals}
\noindent \textbf{Clarify the access to materials}:
As suggested by \cite{Stodden1240}, a digital object identifier (DOI) that uniquely discovers the related computer code and data sets should be assigned by the journal.  Also, they recommend the sharing of the DOIs in trusted open repositories with sufficient detailed information such as the title, authors, version, software description (e.g., inputs, outputs, dependencies, etc.) if possible.  In fact, the journal \textit{Statistics in Medicine} has already followed the publisher \textit{Wiley}'s data-sharing policies by including a DOI but only for data sets (click \href{https://onlinelibrary.wiley.com/page/journal/10970258/homepage/forauthors.html}{\textcolor{blue}{here}} for more information):
\begin{quote}
`Upon acceptance for publication, data files will be deposited to \href{https://figshare.com/}{figshare}, by \textit{Wiley}, on behalf of the authors. The data will be assigned a single DOI and will be automatically and permanently associated with the HTML version of the published manuscript.'
\end{quote}

The publisher, \textit{Wiley}, shows it commitment to a more open research landscape, facilitating faster and more effective research discovery by enabling reproducibility and verification of data, methodology and reporting standards.  They encourage authors of articles published in their journals to share their research data including, but not limited to: raw data, processed data, software, algorithms, protocols, methods, and materials.  According to the introduction on the \textit{Wiley} website (see \href{https://authorservices.wiley.com/author-resources/Journal-Authors/open-access/data-sharing-citation/data-sharing-policy.html}{\textcolor{blue}{here}}), they produce a table in order to understand the various standardized data sharing policy categories which we reproduce in Table \ref{wiley}. 

\begin{table}[htbp]
  \centering
    \begin{tabular}{|p{11em}|p{9em}|p{7em}|p{7em}|}
    \toprule
    \multicolumn{1}{|c|}{} & Data availability statement is published & Data has been shared & Data has been peer reviewed \\
    \midrule
    Encourages Data Sharing & Optional & Optional & Optional \\
    \midrule
    Expects Data Sharing & Required & Optional & Optional \\
    \midrule
    Mandates Data Sharing & Required & Required & Optional \\
    \midrule
    Mandates Data Sharing \& Peer Reviews Data & Required & Required & Required \\
    \bottomrule
    \end{tabular}
      \caption{A table from the publisher \textit{Wiley}'s website to help to understand the various standardized data sharing policy categories.}
  \label{wiley}
\end{table}

From inspecting the papers published in \textit{Statistics in Medicine}, we believe that the journal only adheres to the first standard (first column): a data availability statement confirming the presence or absence of shared data is not necessarily provided.   We do not believe that it adheres to the second standard (second column): if data have been shared in a data repository, the link is not checked to ensure the validity.  It also does not adhere to the third standard (third column): the replicability of linked data is not peer-reviewed.  These are inconsistent with the claim on the journal's website that it `\textbf{expects} that data supporting the results reported in the paper will be archived in an appropriate public repository'.  Even for published papers in this journal which did provide executable computer code files and fMRI data, data was not made available in figshare, and certainly no DOI was issued.  In one case, all the MATLAB computer code and the related data sets were posted on the first author's `Faculty \& Staff' page on the university website for downloading, with researchers requiring additional steps to access the materials.  On the other hand, for those recently published papers containing the `Data Availability Statement' portion (2 papers), data sharing is still not applicable as no data was created nor did they provide their preprocessed version of the data (or the raw data and a preprocessing script).  This case study for \textit{Statistics in Medicine} illustrates the current difficulties in assigning DOIs to data sets.  To the best of our knowledge, no fMRI-related paper in \textit{Statistics in Medicine} mention the data set DOI at this time, which renders the efforts of the journal and publisher unfulfilled.

As a valuable alternative, we recommend \textit{Journal of the Royal Statistical Society: Series C}'s practice for how it arranges the materials of its recently published papers. JRSS,C sorts the majority of the computer code and the related data of its papers on its website in chronological order, although some earlier published papers have missing resources.  This convenient search method helps researchers easily find all the related materials of papers simply by looking up the corresponding volume number. \\

\noindent \textbf{Material availability statement}: 
Inspired by \textit{Biometrics}'s policy (and also mentioned in some of \textit{Wiley}'s standardized data sharing policies):

\begin{quote}
`Authors are required to provide a `Data Availability Statement' to describe the availability or the absence of shared data. Please ensure the main manuscript contains this statement which should be a new, unnumbered section placed immediately before the list of references.' 
\end{quote}

\noindent We strongly recommend that authors create a new section in their articles called `Material Availability Statement' in which they clearly state the availability or the absence of their computer code AND data files.  The new section will ameliorate the inconsistent locations where authors choose to provide information on their files.  In the 93 published papers in our study, information on the the code repository links or the data resource were placed in the abstract, introduction, data description, data analysis, conclusion, and supplementary files, which almost include all sections of the paper.  We believe a unified placement of this information on the availability of computer code and data will be more convenient and helpful to readers and for paper review.  In fact, \textit{Biostatistics} (link \textcolor{blue}{\href{https://academic.oup.com/biostatistics/pages/General_Instructions }{here}}) set up a similar reproducible research policy that states 
\begin{quote}
`... papers in the journal should be kite-marked D if the data on which they are based are freely available, C if the authors' code is freely available, and R if both data and code are available ...' 
\end{quote}

This is highly commendable, the policy was introduced in 2010, but \cite{peng2011} found that by July 2011, only 21 of 125 published papers in \textit{Biostatistics} had a kite-mark, including five articles with an ``R'' kite mark. Even though \textit{Biostatistics} introduced the kite-mark system in 2010, only papers published after 2018 from our data set were actually kite-marked.  From these 13 published papers, 3 articles are C-marked (code available), 1 is D-marked (data available) and 2 are R-marked (both are available). However, articles with the D/R mark may only provide simulated data without sharing the real fMRI data.

\noindent \textbf{Reproducibility check}:  
Although it would be very difficult to achieve in a short amount of time, we hope all statistics journals will regulate reproduction standards for each paper, and (wherever possible) check the reproducibility of its already published papers (this would indicate a real commitment to reproducibility).  Going forward, we recommend that instead of using opaque and optional words like `encourage', `expect' and `should', journals should use stronger words such as `require' and `must', which will raise more awareness of reproducibility.  Along with the requirements, submitted papers (or at least papers that are invited for revisions) should be checked in detail for computer code and data repositories, openly licensed artifacts, reproducibility, and the capacity of independent use by other scholars. It should be considered a necessary task for the reviewers (or the journal should have specific reviewers focused on reproducibility similar to JASA) during the review process.  In addition, the editors of the journal should reserve the right to refuse publication of any paper for which the justification for failing to provide data (or details of how to access data), computer code, or any supporting files for replication is deemed inadequate.

As noted in Section~\ref{subsec:jasa}, as of September 2021, the \textit{Journal of the American Statistical Association} has made considerable changes to its requirements for computer code and data'.   Most significantly, it has stipulated that either one of the reviewers or the  JASA associate editors for reproducibility (AER) will carry out a reproducibility review of the work.  We are hopeful and look forward to seeing the impacts of these changes. 

Finally, given the workload currently on editors and journals, another possibility is the creation of non-profit reanalysis centres attached to respected statistics and biostatistics university departments similar to outreach consultancy groups run by PhD students in statistics and biostatistics departments. 

\noindent \textbf{Supplementary materials}:
While Table \ref{requirements} provides detailed descriptions of the computer code and data requirements for each journal, the journals also provide statements for the submitted supplementary materials/files (here, we only focus on the attached computer code and data in the supplementary materials).  For example, 

\noindent \textit{Biometrics} states that:
\begin{quote}
`Code and data are not subject to a formal review and will be posted ``as-is.'' ';
\end{quote}

\noindent \textit{Journal of Computational and Graphical Statistics} states that:
\begin{quote}
`The supplements are subject to editorial review and approval.'
\end{quote}

\noindent \textit{Journal of the American Statistical Association} states that:
\begin{quote}
`Supplementary files should be supplied for review along with the manuscript at the initial submission.'
\end{quote}

\noindent and \textit{Statistics in Medicine} states that:
\begin{quote}
`The publisher is not responsible for the content or functionality of any supporting information supplied by the authors. Any queries (other than missing content) should be directed to the corresponding author for the article.'
\end{quote}
\noindent The other three journals we consider in this paper do not specifically detail how they deal with the supplementary materials.  While supplementary material policies are not the same as reproducibility policies, a clear statement on the material-review process in turn greatly influences the quality of the attached materials. Therefore to improve reproducibility, we deem a mandatory check on these indispensable materials.  

\noindent \textbf{Rewards for authors and reviewers}:   
Following the proposal from \cite{Stodden1240}, journals can improve their reproducibility by rewarding reviewers who take extra effort to verify computational findings and authors who facilitate such a review. Annual accolades such as prizes and/or badging or even cash could be awarded to both authors and reviewers. In \textit{Biometrics}, Open Research Badges have been applied in recent years in partnership with the non-profit Center for Open Science (COS) to recognize authors' contributions to reproducibility work such as sharing research instruments and materials in a publicly-accessible format, providing sufficient information for researchers to reproduce procedures and analyses.  We strongly recommend that all reviewers working towards reproducible publications should also be considered as candidates for such badges.  Although applying and qualifying for badges is not a requirement for publication, these badges are a further incentive for authors and reviewers to participate in the Open Research Movement and thus to increase the visibility and transparency of the research.  

In general, as researchers, institutions, funders of research, and governments gradually see the benefit of open content, the necessity and urgency of ensuring reproducibility for research will be mentioned more frequently. The field will not change promptly, but developing a culture of reproducibility is paramount and will require time, patience and effort of the community. We hope the work and strength of different groups come together to facilitate reproducibility and make the above effects commonplace in the future.

\subsection{Strengths and limitations}
In this paper, we examine reproducibility in the field of statistics by attempting to reproduce the results in 93 published papers in prominent journals utilizing functional magnetic resonance imaging (fMRI) data during the 2010-2021 period.  While there is currently a great many research papers on reproducibility in other computational
fields, this is the first study on the reproducibility of results in statistics.  Overall, we conclude that while several journals have good policies in terms of reproducibility, the current reproducibility statistics are poor and the level needs to improve. We detail the reasons for the low reproducibility numbers and provide  author-specific and journal-specific recommendations to improve the research reproducibility in statistics.

In this work, we focus on the reproducibility of research papers in the field of statistics based on fMRI.  The reason for the spotlight on fMRI is that we understand fMRI very well and its idiosyncrasies.  However, we understand that our results might not generalize to every area of application.  This may be due to some particular properties of fMRI data.  For example, first, fMRI data can be shared in various formats such as raw data, data that has been somewhat preprocessed, region of interest (ROI) time series data or the complete preprocessed data set.  
While raw fMRI data is preferable to providing no data, it puts the onus on the researcher attempting to replicate the results to preprocess the fMRI data.  As we discuss earlier, it is extremely difficult to obtain precisely the same preprocessed data from the raw data as there is not an established sequence in carrying out the preprocessing steps. Second, fMRI data is expensive to obtain, which means that many neuroscientists are unwilling to make it available openly until the data has been exhausted in terms of creation of research papers.  However, there are many open source fMRI data sets available to statisticians such as \url{openfmri.org} and \url{http://www.humanconnectomeproject.org/}.  Hence, fMRI lies between data that is very open (e.g., stock prices and returns on the main indices) and data that is truly proprietary.

Another limitation of the work is that we did not contact the authors of the published papers for their computer code and data.   We took this step as there is no quality control when the authors share the computer code and data.  Hence, we believe the reproducibility standards should be set at the journal level to maintain standards and  shift expectations for transparency.  However, anecdotally, in our experience, statisticians are very good at sharing computer code and data when requested by email.  
Finally, while the statistical analysis of data is important in any study, there are other vital steps in research, and there is much more to statistical analysis of data than checking the calculations. It is essential to recognize that reproducibility also requires the understanding of the substantive question.  Going forward, statistical research (and the reproduction of this research) should be mindful of this and consider the substantive context adequately.

\newpage

\noindent \LARGE \textbf{Appendix}

\noindent \large \textbf{Computer code details}

\normalsize

%\begin{longtable}[htbp]{cccp{5em}cp{12em}}
\footnotesize
\begin{longtable}[htbp]{|p{1cm}p{2.3cm}p{1.7cm}p{3cm}p{1.3cm}p{5cm}|}
    \hline \hline
    Journal & Software & Data type & Data Note & Working code & Code errors \\
    \hline
    AOAS & R package & PP & & NA & Only R package is provided. \\
	\hline
    AOAS & R package & (Partly)PP & 30/1250 voxels & NA & Only R package is provided. \\
	\hline
    AOAS & Matlab & PP &  & N & A key brain file 'n33\_buckner17\_k286.mat' is not provided. \\
	\hline
    AOAS & Matlab & PP &  & Y & Results differ from the paper (significant ROI-SNP connection: 24/26) \\
	\hline
    AOAS & R package & PP &  & NA & Only R package is provided. \\
	\hline
    AOAS  & Matlab & Sim &  & Y &  \\
	\hline
	Biom & Matlab & PP & Meta-data & Y &  \\
	\hline
    Biom  & R & PP &  & Y &  \\
	\hline
    Biom & Matlab & Sim &  & Y &  \\
	\hline
    Biom & Matlab & Sim &  & N & A key brain file brain\_sample.map was not provided. \\
	\hline
    Biom  & R & Sim &  & Y &   \\
	\hline
    Biom & R & Sim &  & Y &  \\
	\hline
    Biom & R package & Raw &  & NA & Only R package is provided.  \\
	\hline
	Biom & R package & N &  & NA & Only R package is provided.  \\
	\hline	
	%!!new
    Biom & R package & Sim &  & Y &   \\
	\hline
	Biom & Matlab & PP &  & Y &   \\
	\hline
	Biom & R package & Sim &  & Y &   \\
	\hline
	Biom & R & PP &  & Y &   \\
	\hline	
	Biom & R package & PP &  & Y &   \\
	\hline		
	Biom & R & N &  & NA & Only include R functions.
   \\
	\hline		

   Bios  & Matlab & PP & & Y &  \\
	\hline
    Bios  & R & (Partly) PP & & N & Error in Wstat[which(Wtrue == 0, arr.ind = TRUE)] : \newline{}  subscript out of bounds\newline{} \\
	\hline
    Bios  & R & (Partly) PP & 50/197 time points. & Y &  \\
	\hline
    Bios  & R & Sim &  & Y &  \\
   	\hline
    Bios  & R & Sim &  & N &file14.Rmd: Error in dlda(x = train.noise, y = train.labels): could not find function "dlda"  \\
	\hline
    Bios  & R package & Sim & & NA & Only R package provided. \\
	\hline
    Bios & R & N & NA & N & No real data provided.  \\
	\hline
	
	%!!NEW
    Bios & R & Sim &  & Y &   \\
	\hline	
	
    JCGS & Python & PP & Preprocessed in the code & Y &  \\
	\hline
    JCGS & R & PP &  & Y &  \\
	\hline
    JCGS & Matlab & Sim &  & Y &  \\
	\hline
    JCGS & R & N & Upon request & N & One key dataset "suicide.rda" is missing. \\
	\hline
	JASA & R & Sim &  & Y &  \\
	\hline
    JASA & R & Sim &  & Y &  \\
	\hline
    JASA & Python+R & Sim &  & N & Can't install smoothfdr package fatal error C1083: Cannot open include file: 'bayes\_gfl.h': No such file or directory \\
	\hline
    JASA & R & Sim &  & Y &  \\
	\hline
    JASA & R package & Raw &  & NA & Only R package is provided.  \\
	\hline
    JASA & Matlab & Raw & Difficulty downloading the data & N & Unable to run without data. \\
	\hline
    JASA & R+Matlab & Raw & Difficulty downloading the data & N & Unable to run without data. \\
	\hline
	
	%!!new
	JASA & R & N &  & NA & Only R functions are provided.  \\
	\hline
	JASA & Matlab & N & Very detailed documentation, but not as organized in the real zipped file. & NA & A key file is missing.  \\
	\hline
	
    JRSS,C & R & (Partly) PP & 1/45 subjects & Y &  \\
 	\hline
    JRSS,C & Linux & PP & meta-data & N & The log file gave errors, for example, make: nvcc: Command not found; Makefile:11: recipe for target 'functions.o' failed\newline{}make: *** [functions.o] Error 127  \\
	\hline
    JRSS,C & R+Matlab+Linux & PP &  & Y &  \\
 	\hline
 	%new!!
    JRSS,C & R & Sim & & Y &  \\
 	\hline 	
 	
    Stat Med & Matlab & PP &  & Y &  \\
	\hline
    Stat Med & R function & N &  & NA & Only R functions are provided. \\
	\hline
	
	\caption{More details on the 47 out of the 93 published papers that provide computer code.  All the papers contained functional magnetic resonance imaging (fMRI) data and were published in seven prominent statistical journals: the \textit{Annals of Applied Statistics} (AOAS), 
\textit{Biometrics} (Biom), \textit{Biostatistics} (Bios),  the \textit{Journal of Computational and Graphical Statistics} (JCGS), the \textit{Journal of the American Statistical Association} (JASA), the \textit{Journal of the Royal Statistical Society: Series C} (JRSS, C) and \textit{Statistics in Medicine} (Stat Med) from 2010 to 2021. PP and (Partly) denote preprocessed data and data preprocessed to some extent, respectively. }

  \label{resultofcode}%
  
\end{longtable}%

\newpage

\bibliography{reproducibility_arxiv}

\end{document}